\documentstyle[11pt,epsfig]{article}
\newcommand \be  {\begin{equation}}
\newcommand \bea {\begin{eqnarray}}
\newcommand \ee  {\end{equation}}
\newcommand \eea {\end{eqnarray}}
\begin{document}

\title{{\bf Have your cake and eat it too: \\ increasing returns
while lowering large risks!}\thanks{We
are grateful to F.~Lacan, E. Malherbes and V. Pisarenko for helpful discussions.}}

\author{\bf J.V. Andersen\thanks{Nordic Institute for Theoretical Physics,
 Blegdamsvej 17, DK-2100 Copenhagen, Denmark}
 ~and~ D. Sornette\thanks{Institute of Geophysics
and Planetary Physics and Department of Earth and Space Science,
University of California, Los Angeles, California 90095 and
Laboratoire de Physique de la Mati\`ere Condens\'ee, CNRS UMR6622 and
Universit\'e des
Sciences, B.P. 70, Parc Valrose, 06108 Nice Cedex 2, France, E-mail:
sornette@cyclop.ess.ucla.edu}
}

\date{University of California, Los Angeles \ \\ \ \\ {\normalsize First
Version: March 1999 \ \\ This Version: March 1999}}

\maketitle

\vskip 2cm
\begin{abstract}

\noindent  Based on a faithful
representation of the heavy tail multivariate distribution of asset returns
introduced previously (Sornette et al., 1998, 1999) that we extend
to the case of asymmetric return distributions, we generalize the return-risk
efficient frontier concept to incorporate the dimensions of large risks embedded
in the tail of the asset distributions. We demonstrate 
that it is often possible to {\it increase} the portfolio return
while {\it decreasing} the large risks as quantified by the fourth and higher
order cumulants. Exact theoretical formulas are validated by empirical tests.

\end{abstract}

\thispagestyle{empty}

\pagenumbering{arabic}

\newpage

\section{Introduction}

One of the most fundamental tenet of economic theory and practice is that returns above
the so-called riskless rate come with increased risks. This is the basis of
Markovitz's portfolio theory (e.g. Markovitz, 1959) and of the CAPM (e.g. Merton, 1990).
Reciprocally, investors want to be compensated for taking risk, that is, they want to earn
a return high enough to make them comfortable with the level of risk they are assuming.
It is thus a fundamental premise of efficient markets
that ``one cannot have both the cake and eat it too'', i.e. 
one cannot increase the return and lower the risk at the same time.
This result stems simply from the linear (resp. quadratic) dependence of the 
average return (respectively variance) of
a portfolio return on the weights of its constituting assets leading to a parabolic
efficient frontier in the return-risk diagram.

In the real world, the variance of portfolio returns provide only a limited 
quantification of incurred risks, as the distributions of returns have ``fat tails''
(e.g. Lux, 1996, Gopikrishnan et al., 1998, Lux and Marchesi, 1999) and 
the dependences between assets are only imperfectly accounted for by the correlation
matrix (e.g. Litterman and Winkelmann, 1998). Value-at-Risk (e.g. Jorion, 1997)
and other measures of risks (e.g. Artzner et al., 1996,
Sornette, 1998, Bouchaud et al., 1998, 
Sornette et al., 1998, 1999) have been developed to account for the larger moves
allowed by non-Gaussian distributions.

Here, we generalize our previously introduced
representation of the heavy tail multivariate distribution of asset returns
(Sornette et al., 1998, 1999)
to the case of asymmetric return distributions. We calculate theoretically and
test empirically the cumulants of a portfolio and generalize the return-risk
efficient frontier concept to incorporate the dimensions of large risks embedded
in the tail of the asset distributions. We demonstrate the novel
remarkable result that it is often possible to improve on the optimal mean-variance
portfolio by {\it increasing} the return
while {\it decreasing} the large risks quantified by the fourth and higher
order cumulants. This is related to and generalizes our previous rigorous result
(Sornette et al., 1998, 1999)
that minimizing the variance, i.e. the relatively ``small'' risks, 
often increases larger risks as measured
by higher normalized cumulants and the Value-at-risk. Thus, putting the emphasis
on the risk quantified by the volatility can be both misleading because 
large risks are still looming and in addition damage profitability.

\section{The asymmetric modified Weibull distribution}

In order to make our approach concrete, 
we assume that price returns $\delta x$ are distributed according to the
following
probability distribution function (pdf)
\bea
P(\delta x)  &=& {Q \over \sqrt{\pi}}
{\gamma_+ \over \chi_+^{\gamma_+ / 2}}~|\delta x|^{{\gamma_+ \over 2}-1} ~~
\exp\left(-\left({|\delta x| \over \chi_+}\right)^{\gamma_+} \right)~~~~~{\rm for}~~
0< \delta x ~, \label{aer}\\
&=& {1-Q \over \sqrt{\pi}}
{\gamma_- \over \chi_-^{\gamma_- / 2}}~|\delta x|^{{\gamma_- \over 2}-1} ~~
\exp\left(-\left({|\delta x| \over \chi_-}\right)^{\gamma_-} \right)~~~~~
{\rm for}~~\delta x < 0~. \label{aera}
\eea
$Q$ is the probability for observing a positive return, the $\chi$'s are the
characteristic returns and the exponent $\gamma$'s control the fatness of the pdf tails,
which can be different for positive and negative returns.

For $Q=1/2$, $\chi_+=\chi_-$ and $\gamma_+=\gamma_-$, we recover the symmetric modified
Weibull pdf studied by Sornette et al. (1998, 1999) and the special case
$\gamma_+=\gamma_-=2$ recovers the standard normal law.
The case when the exponents $\gamma$ are smaller than one
corresponds to a ``stretched'' exponential with a tail fatter
than an exponential and thus much fatter than a Gaussian, but still thinner
than a power law. Stretched exponential pdf's have been found to provide a
parsimonious and
accurate fit to the full range of currency price variations at daily
intermediate time scales (Laherr\`ere and Sornette, 1998). This
stretched exponential model is also validated theoretically by the recent
demonstration that
the tail of pdf's of products of a finite number of random variables
is generically a stretched exponential (Frisch and Sornette, 1997), in
which the exponent $\gamma$ is
proportional to the inverse of the number of generations
(or products) in a multiplicative process.

\section{Nonlinear change of variable}

Let us pose
\bea
y_+ &=& (\delta x)^{\gamma_+/2}~~~~~ {\rm for}~\delta x > 0~, \label{aetyu}\\
y_- &=& - |\delta x|^{\gamma_-/2}~~~~~ {\rm for}~\delta x < 0~.  \label{jllllll}
\eea
Inversely, we have
\bea
\delta x &=& y_+^{q_+}~~~~~~~ {\rm for}~\delta x > 0~,~~~{\rm with}~~q_+
\equiv {2 \over \gamma_+}~,
\label{hkmqmmq}\\
\delta x &=& -|y_-|^{q_-}~~~{\rm for}~\delta x < 0~,~~~{\rm with}~~q_-
\equiv {2 \over \gamma_-}~.
\label{jfmmaqmmq}
\eea

The change of variable (\ref{aetyu},\ref{jllllll})
from $\delta x$ to $y$ leads to a Gaussian pdf for the
$y$-variable defined in each semi-infinite domain:
\bea
P(y_+) &=& {2Q \over \sqrt{2\pi}~\sigma_+}~\exp\left(-{y_+^2 \over 2
\sigma_+^2}\right)~,
~~~~~{\rm where}~\sigma_+^2 = {1 \over 2} \chi_+^{\gamma_+}~, \label{jkmmmqmmq}\\
P(y_-) &=& {2(1-Q) \over \sqrt{2\pi}~\sigma_-}~\exp\left(-{y_-^2 \over 2
\sigma_-^2}\right)~,
~~~~~{\rm where}~\sigma_-^2 = {1 \over 2} \chi_-^{\gamma_-}~. \label{kqljjnvmq}
\eea

Using a maximization entropy principle, one can then show 
(Sornette et al., 1998, 1999) that the correlations between
the $y$ variables of different assets provide the most efficient and parsimonious
multivariable representation.
This transformation has also been used
for the analysis of particle physics experiments (Karlen, 1998) and much earlier
for the treatment of bivariate gamma distributions (Moran, 1969). It can also be
viewed as a concrete implementation of the copula representation of 
dependence between assets (e.g. Embrechts et al., 1998, 1999). Generalizations
to other non-Gaussian pdf's are discussed in Sornette et al. (1999).

We have made empirical tests on three assets, using annualized daily returns of stock prices  
of Chevron (CHV) and Exxon (XON) in the period Jan. 1970 - Mar. 1999, and of the
Malaysian Ringit (MYR) against the US dollar
in the period Jan. 1971 - Oct. 1998. The CHV-XON pair is 
among the most strongly connected group of stocks in the S\&P 500 index
while the Malaysian Ringit is essentially uncorrelated to the Chevron and Exxon stocks.
These extreme cases allow us to test the influence of correlations.
Especially for strongly correlated stocks, we  
have shown (Sornette et al., 1998) that a change of variable like 
Eq.~(\ref{aetyu},\ref{jllllll}) leads to a covariance matrix which is much more stable 
compared to the usual covariance matrix. 

Fig.~\ref{fig1} shows in a $\log$-$\log$ plot the $y(r)$ transformation 
(\ref{aetyu},\ref{jllllll}) calculated 
from the empirical
positive and negative returns of the Chevron and Exxon stocks and for the
Malaysian Ringgit against the US dollar (MYR). Assuming that price returns are distributed 
according to an asymmetric modified 
Weibull (\ref{aer},\ref{aera}), the slope of the $y(r)$-plot gives for 
large $|r|$-values the exponents $\gamma_+/2$ and $\gamma_-/2$. 
The positive and negative returns of each asset 
are seen 
to have almost the same slope for large $r$ values, and consequently we will assume  
for each asset that $\gamma_+=\gamma_- \equiv \gamma$ in the sequel. 
The linearity of the $y(r)$ plots 
for large $r$ values show that the large tails of the pdf's are indeed to a very good 
approximation distributed according to a modified Weibull distribution Eq.~(\ref{aer},\ref{aera}),
with $\gamma \approx 1.4$ (CHV), $\gamma \approx 1.2$ (XON) and $\gamma \approx 0.62$ (MYR).
For small and intermediate $r$ values,
the $y(r)$ curves have a slope close to 1 (indicated by the $y=r$ line), which means that small 
and intermediate returns are distributed according to a Gaussian distribution. Because of the 
finite resolution of the data (the data has a lower bound for the return), $y(r)$ approaches a 
constant value for the smallest values of $r$.

\section{Portfolio theory for the diagonal case}

In this short letter, we present the theory for the diagonal case where assets
are uncorrelated. This is already sufficient to illustrate the most important
results. Especially in the case of fat tails (exponents $c <1$), correlations are less important
than a precise determination of the tails (Sornette et al., 1998). We will however
present some empirical tests with uncorrelated and with correlated assets, in 
order to illustrate the importance of correlations. Sornette et al. (1999) treat
the case of correlated assets with symmetric distributions with the same
exponent $\gamma$. Generalization to the asymmetric case and with different exponents
$\gamma$ will be reported elsewhere.

The discrete time estimation of the returns $\delta x_i(t)$
are  $\delta x_i(t) \equiv 
\delta p_i(t) /p_i(t) =(p_{i}(t+1) - p_i(t))/p_i(t)$, where
$p_i(t)$ is the price of asset $i$ at time $t$.
The total variation of the value of the portfolio made of $N$ assets between
time $t-1$ and $t$ reads
\be
\delta S(t) = \sum_{i=1}^N W_i \delta p_i(t) = \sum_{i=1}^N w_i \delta x_i(t)~,
\label{sumport}
\ee
where $W_i$ is the number of shares invested in asset $i$ and
$w_i = W_i p_i$ is the weight in capital invested in the $i$th asset
at time $t$ in the portfolio. We will assume normalization, i.e.
$\sum_{i=1}^N w_i = 1$,
thus leading to a dynamical reallocation of the assets in the portfolio.

The expression (\ref{sumport}) can be expressed in terms of the variables
$y_i$'s defined by (\ref{aetyu},\ref{jllllll}) as follows
\be
\delta S(t) = \sum_{i=1}^N w_i ~\epsilon_i |y_{\epsilon_i}|^{q_{\epsilon_i}}~,
\label{sumpqgqqgort}
\ee
where $\epsilon_i$ is the sign of $\delta x_i$.
All the properties of the portfolio are contained in
the probability distribution $P_S(\delta S(t))$ of $\delta S(t)$. We would
thus like
to characterize it,
knowing the multivariate distribution of the $\delta x_i$'s (or
equivalently the multivariate
Gaussian distribution of the $y_i$'s) for the different assets. The
general formal solution reads
\be
P_S(\delta S) = C \prod_{i=1}^N \biggl( \int dy_i \biggl) ~ e^{-{1 \over 2}
~{\bf y}' V^{-1} {\bf y}}
~\delta\biggl(\delta S(t) - \sum_{i=1}^N w_i
\epsilon_i |y_{\epsilon_i}|^{q_{\epsilon_i}} \biggl) ~ .
\label{soqglut}
\ee
Taking the Fourier transform
$\hat P_S(k) \equiv \int_{-\infty}^{+\infty} d\delta S ~P_S(\delta S)~
e^{-ik \delta S}$ of
(\ref{soqglut}) gives
\be
{\hat P}_S(k) = \prod_{i=1}^N \biggl( \int dy_i \biggl) ~ e^{-{1 \over 2} ~
{\bf y}' V^{-1} {\bf y} +
ik~\sum_{i=1}^N ~w_i ~\epsilon_i |y_{\epsilon_i}|^{q_{\epsilon_i}}} ~ .
\label{soqqgglut}
\ee

Using the explicit expression of 
the form of the distributions (\ref{jkmmmqmmq},\ref{kqljjnvmq}), we get
\bea
{\hat P}_S(k) &=& \prod_{i=1}^N \biggl[ 2(1-Q_i) \int_{-\infty}^{0}
{dy_i \over \sqrt{2\pi}~\sigma_{i-}}~\exp\left(-{y_i^2 \over 2 \sigma_{i-}^2}
- ik w_i |y_i|^{q_{i-}}\right) \nonumber \\
&+& 2Q_i \int_{0}^{+\infty}{dy_i \over \sqrt{2\pi}~\sigma_{i+}}~
\exp\left(-{y_i^2 \over 2 \sigma_{i+}^2} + ik w_i y_i^{q_{i+}}\right)\biggl] ~.
\label{soqqgqqglurft}
\eea
Expanding the exponential $\exp \left(ik w_i |y_i|^{q_{i}}\right)$ in powers of
its argument, we get
\be
{\hat P}_S(k) = 2 \prod_{i=1}^N \left[ \sum_{m=0}^{+\infty} {\left(ikw_i
\right)^m
\over m!}~
\left( (-1)^m (1-Q_i) \sigma_{i-}^{mq_{i-}} \langle y^{mq_{i-}} \rangle_+
+ Q_i \sigma_{i+}^{mq_{i+}} \langle y^{mq_{i+}} \rangle_+ \right) \right]~,
\label{dqddqgh}
\ee
where
\be
\langle y^{\alpha} \rangle_+  \equiv \int_0^{+\infty} {dy \over
\sqrt{2\pi}}~y^{\alpha}~e^{-{y^2 \over 2}}
= {2^{{\alpha \over 2}-1} \over \sqrt{\pi}}~\Gamma \left({\alpha \over 2}+{1 \over
2}\right)~,
\ee
and $\Gamma$ is the Gamma function. Replacing in (\ref{dqddqgh}), we obtain
\be
{\hat P}_S(k) = \prod_{i=1}^N \left[ \sum_{m=0}^{+\infty} {\left( ikw_i
\right)^m
\over m! } ~M_i(m) \right]~,
\label{dqdqqdqghhh}
\ee
where
\be
M_i(m) = {1 \over \sqrt{\pi}} \left( (-1)^m (1-Q_i) 2^{mq_{i-}/2}
\sigma_{i-}^{mq_{i-}}
\Gamma\left({mq_{i-} \over 2}+{1\over 2}\right)
+ Q_i 2^{mq_{i+}/2} \sigma_{i+}^{mq_{i+}} \Gamma\left({mq_{i+}\over
2}+{1\over 2}\right) \right)~.
\ee

For symmetric distributions with $q_{i+}=q_{i-}$, i.e. $\gamma_{i+}=\gamma_{i-}$,
$\sigma_{i+}=\sigma_{i-}$ and $Q_i = 1/2$, we retrieve our previous result
(Sornette et al., 1999) that all the odd order terms in the sum over $m$
cancel out\,:
\be
{\hat P}_S(k) = \prod_{i=1}^N \left[ \sum_{n=0}^{+\infty} {\left( ikw_i
\right)^{2n}
\over (2n)! \sqrt{\pi}} ~ 2^{nq_i}~\Gamma\left(nq_i+{1 \over 2}\right)
\sigma_i^{2nq_i} \right]~.
\ee

The expression $\sum_{m=0}^{+\infty} {\left( ikw_i \right)^m
\over m! } ~M_i(m)$ in (\ref{dqdqqdqghhh}) is similar to the expansion of a
characteristic
function in terms of moments. We need to get the corresponding expansion in
terms
of cumulants, i.e. find the coefficients $c_n$ such that
\be
\sum_{m=0}^{+\infty} {\left( ikw_i \right)^m \over m! } ~M_i(m) =
\exp \left( \sum_{n=1}^{+\infty} {\left( ik \right)^m
\over n! } ~c_i(n) \right)~.
\ee
By identifying the same powers of $k$ term by term, we get the cumulants.
Then, using the product in (\ref{dqdqqdqghhh}) of the exponentials from
$i=1$ to $N$, we obtain
the cumulants of the portfolio distribution as
\bea
c_1 &=& \sum_{i=1}^N w_i M_i(1) ~, \label{c1}\\
c_2 &=& \sum_{i=1}^N w_i^2 \left(M_i(2) - M_i(1)^2\right) ~, \label{c2} \\
c_3 &=& \sum_{i=1}^N w_i^3 \left( M_i(3) -3 M_i(1)M_i(2) + 2M_i(1)^3\right)
~, \label{c3}\\
c_4 &=& \sum_{i=1}^N w_i^4 \left( M_i(4) -3 M_i(2)^2 -4 M_i(1) M_i(3)
 + 12 M_i(1)^2 M_i(2) - 6 M_i(1)^4\right)   ~, \label{c4} \\
c_5 &=& \sum_{i=1}^N w_i^5 \biggl(M_i(5)  - 5M_i(4) M_i(1)  -
10 M_i(3)  M_i(2)  +20 M_i(3)  M_i(1) ^2 + 30 M_i(2)^2 M_i(1) \nonumber\\
&& -60M_i(2) M_i(1) ^3+24 M_i(1) ^5\biggl)~  ,\label{f1e}\\
c_6 &=& \sum_{i=1}^N w_i^6 \biggl(M_i(6)  - 6M_i(5)M_i(1)  -15
M_i(4)M_i(2)  +30 M_i(4)M_i(1)^2 -10 M_i(3)^2 \nonumber\\
&&+120 M_i(3) M_i(1) M_i(1)  -120 M_i(3) M_i(1)^3 +30 M_i(2)^3\nonumber\\
&& -270 M_i(2)^2 M_i(1)^2 +360 M_i(2) M_i(1)^4 -120 M_i(1)^6\biggl) ~ \label{c6} .
\eea
Higher order cumulants are obtained by using the formulas given for instance by
Stuart and Ord (1994).
The first cumulant $c_1$ provides the average gain $\langle \delta S \rangle$
and the second cumulant $c_2$ is the variance of the portfolio gain. The higher order
cumulants as well as the excess kurtosis $\kappa \equiv c_4/c_2^2$
quantify larger risks occurring with smaller probabilities but larger impact.

Fig.~\ref{fig2} presents a comparison of the empirical determined $c_n$'s and those
determined from the equations (\ref{c1}-\ref{c4}), for a portfolio constituted of
the Malaysian Ringgit (MYR) and the Chevron stock (CHV). This choice is made because
MYR is essential uncorrelated to CHV and the above calculation should thus apply directly.
For an extension of the theory to correlated assets, see Sornette et al. (1999). 
To perform the empirical test shown in figure~\ref{fig2},
we first determined the exponents $\gamma_+=\gamma_- \equiv \gamma$ 
from a regression of the linear parts of
the $y(r)$ functions for large values of $|r|$ shown in figure 1. 
We then use these $\gamma$'s to estimate 
the coefficients $\chi_{i+}, \chi_{i-}$ from the empirical averages
$\chi_{i\pm} = \langle (\delta x_{\pm})^{\gamma_{i}}\rangle_\pm$. 
The notation $\langle ~ ~ \rangle_\pm $ represents an average taken with respect 
to positive/negative returns of the data. The asymmetric weight parameter $Q_i$ is determined 
from the asset $i$ as the ratio of the number of positive returns over the total number of
returns. The error bars shown in the figure are 
determined from the observation that the main source of error comes from a mispecification 
of the tail exponent $\gamma$'s and we assume conservatively an error of $\pm 0.05$ on the 
$\gamma$ values.
Fig.~\ref{fig2} shows a very good agreement between theory and the direct empirical determination 
of the cumulants. There is some discrepancy for the third order
cumulant $c_3$, which reflects our simplification to use symmetric tails with 
$\gamma_+=\gamma_- \equiv \gamma$ in our calculations Eq.~(\ref{c1}-\ref{c4}). As a consequence,
the sole contribution to the odd-order cummulants stems from the difference between
$\chi_{i+}$ and $\chi_{i-}$ and between $Q_i$ and $1/2$. An
additional asymmetry in the shape of the tail captured by $\gamma_+ \neq \gamma_-$, however 
small, can easily make the agreement adequate between the theoretical and empirical $c_3$. We have 
chosen not to incorporate this additional complexity in order to keep the number of degrees 
of freedom as small as possible. The 
even-order cumulants and the excess kurtosis $\kappa$ are much less sensitive to the asymmetry 
in the exponents $\gamma_+,\gamma_-$.

The portfolio with minimum variance $c_2$ has the optimal weight $w_1 = 9.5\%$, where
the index $1$ stands for the Chevron stock, i.e. the weight $w_2 =1-w_1$ of the Malaysian Ringgit
is $90.5\%$. In comparison, the portfolio with minimum fourth cumulant has an investment ratio
of $w_1=38\%$ in Chevron and $w_2 =62\%$ in the Malaysian Ringgit. It is clear that
the minimum variance portfolio has a rather large fourth cumulant, i.e. minimizing the small
risks quantified by the second order cumulant comes at the cost of imcreasing the largest risks
quantified by the fourth order cumulant (Sornette et al., 1998, 1999).

Fig.~\ref{fig3} illustrates another even more interesting phenomenon.
We compare the daily returns and the cumulative wealth of two portfolios.
The first $c_1-c_2$ portfolio has a minimum variance $c_2$ (Chevron weight $w_1=0.095$ and
Malaysian Ringgit weight $w_2 = 0.905$). The second $c_1-c_4$ portfolio has a  minimum fourth-order
cumulant (Chevron weight $w_1=0.38$ and
Malaysian Ringgit weight $w_2 = 0.62$). The horizontal dotted lines  
in the daily return plots are the maximum values sampled
for the returns of the $c_1-c_4$ portfolio. Notice 
that the daily returns of the minimum variance portfolio exceeds these bounds. This
illustrates vividly that, while most of the time the fluctuation of the returns are smaller
for the $c_1-c_2$ portfolio, fluctuations with larger amplitudes
and thus larger risks are observed in this minimum variance
portfolio: again, minimizing small risks can lead to a dangerous increase of large risks
(Sornette et al., 1998, 1999).
Furthermore, the cumulative wealth of the $c_1-c_2$ portfolio with $w_1=0.095$ is 
drastically inferior to that accrued in the $c_1-c_4$ portfolio with $w_1=0.38$. 
In other words, you can have your cake and eat it too: decrease the large risk (those
that count for the safety of investment houses and for regulatory agencies) and 
increase the profit! This example illustrates how misleading can be the focus on the 
variance as a suitable measure of risks and how limited is the use of standard portfolio
optimization techniques. Not only they do not provide a suitable 
quantification of the really dangerous market moves, in addition they miss 
important profit opportunities.

Fig.~\ref{fig4} is the same as Fig.~\ref{fig2} for a portfolio constituted 
of the Exxon and the Chevron stocks.
Due to the very large correlation between the two assets, the departure between
theory and experiments is a measure of the importance of correlations that have
been neglected in the above formulas, expecially in this 
case where the exponents $\gamma$ for the pdfs of the two stocks
are relatively large around $1.4$ and $1.2$ respectively, i.e. the pdf tails
are relatively ``thin''. This constitutes a worst-case scenerio for the application of
the above theory that is best justified for exponents $\gamma<1$ (recall that the standard
Gaussian regime corresponds to $\gamma=2$). 
Nothwithstanding this limitation, the results conform qualitatively to 
our previous discussion: the best variance gives a substantially larger risk for large moves and
the return is sub-optimal.

\section{Efficient Portfolio Frontiers}

Based on our previous calculation, it is straightforward to construct
the optimal mean-variance portfolios from the knowledge of the cumulants $c_1$ and $c_2$
as a function of the asset weights $w_i$. Similarly, we introduce the optimal $c_1-c_4$ 
portfolios. 

For a given mean return $c_1$, the portfolios that minimize the risks expressed 
through $c_2$ given by Eq.~(\ref{c2}) or by $c_4$ given by
Eq.~(\ref{c4}))) are determined from the conditions
\bea
{\partial \over \partial \omega_j} \left[c_2 - \lambda_1 c_1 - \lambda_2 
\sum_i \omega_i  \right]_{|{\omega_j=\omega^*_j}}  &=& 0 ~,  \\
{\partial \over \partial \omega_j} \left[c_4 - \lambda_1 c_1 - \lambda_2 
\sum_i \omega_i  \right]_{|{\omega_j=\omega^*_j}} &=& 0 ~,
\label{optimalconditions}
\eea
where the $\omega^*_j$ denote the weights for an optimal portfolio. From the normalization
condition 
\be
\sum_i \omega_i  = 1 ~,  
\label{sumomega}
\ee
one of the Lagrange multipliers among $\lambda_1, \lambda_2$ can be eliminated.  
Let us define $cn$ such that the expressions (\ref{c1},\ref{c2},\ref{c4}) read
\bea
c_1 & \equiv &\sum_i \omega_i  c1_i ~,  \\
c_2 & \equiv &\sum_i \omega_i^2  c2_i ~,  \\
c_4 & \equiv &\sum_i \omega_i^4  c4_i ~. 
\label{cn}
\eea
The efficient frontier for the mean-variance $c_1-c_2$ porfolios is given by: 
\bea
c_1 & =&  {1 \over 2 \lambda_1} (A - B^2/D) + B/D ~,  \\
c_2 & =&  {1 \over 4 \lambda_1^2} (A - B^2/D) + 1/D ~,  ~~~~~{\rm with}\\
A & \equiv & \sum_i {c1_i^2 \over c2_i} ~,  \\
B & \equiv & \sum_i {c1_i \over c2_i} ~, \\
D & \equiv & \sum_i {1 \over c2_i} ~. \\
\label{c1-c2}
\eea
Varying $\lambda_1$ then traces out the efficient frontier. 
Likewise the efficient frontier for the $c_1-c_4$ portfolios is given by: 
\bea
c_1 & \equiv & \sum_i \omega^*_i  c1_i ~,  \\
c_4 & \equiv & \sum_i (\omega^*_i)^4  c4_i ~,~~~~~~{\rm with}  \\
\omega^*_i & = & {1 \over \sum_i \pm |(c1_i - \lambda_2)/(4c4_i)|^{1/3}}
\pm |{(c1_j - \lambda_2) \over 4c4_j}|^{4/3} ~, 
\label{c1-c4}
\eea
with $+$ if $c1_j > \lambda_2$ and $-$ otherwise. 

\begin{figure}
\epsfxsize=15cm
\epsfbox{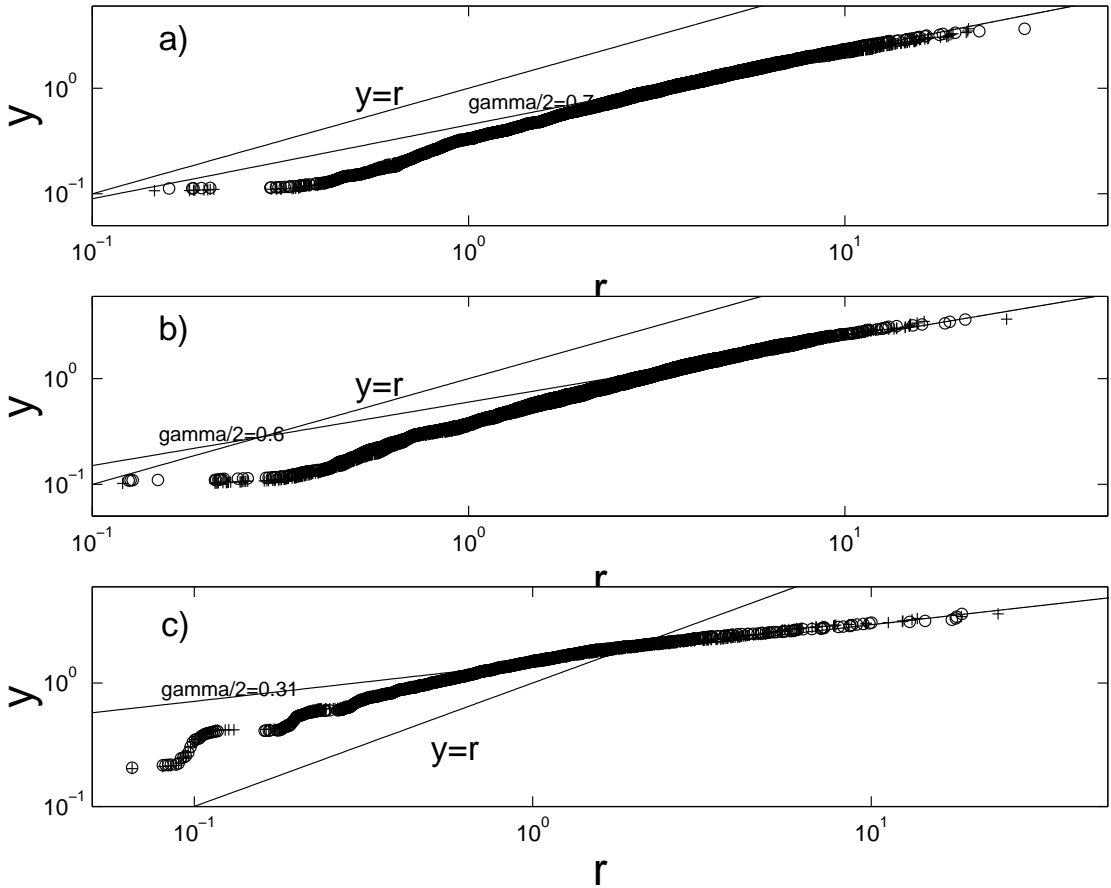}
\caption{ $y(r)$-transformation defined by equations (\ref{aetyu},\ref{jllllll})
for the period from january 1971 to oct. 1998. $+$ corresponds to 
positive returns and $o$ to negative returns. The daily returns $r$ are expressed
in annualized percentage. a) Chevron stock (CHV), b) Exxon stock (XON), 
c) Malaysian Ringgit against US dollar (MYR).}
\label{fig1}
\end{figure}

\begin{figure}
\epsfxsize=15cm
\epsfbox{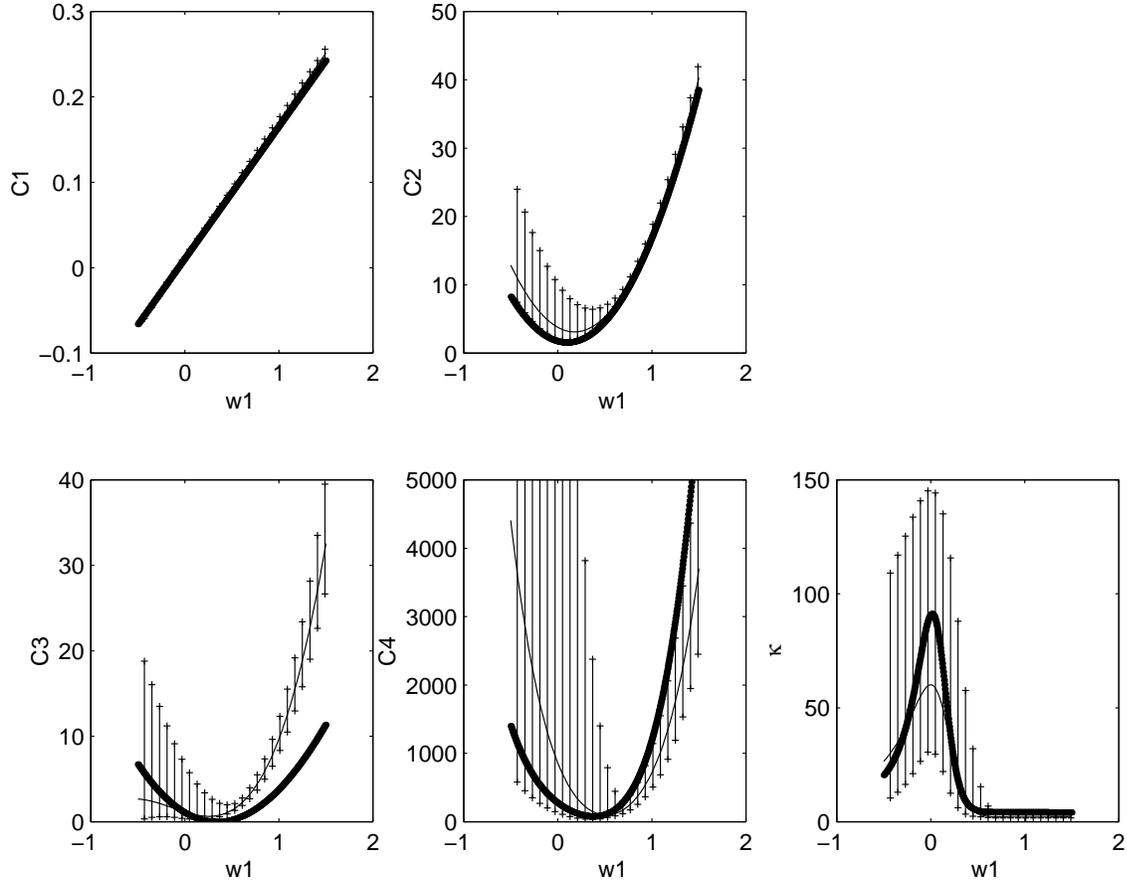}
\caption{Comparison of the empirically determined cumulants $c_n$ and excess kurtosis 
$\kappa$ (fat solid line) to the theory Eq.~(\ref{c1}-\ref{c4}) 
(thin solid line) using the exponents $\gamma_i$ 
determine from Fig.~\ref{fig1} for a portfolio constituted of the Malaysian Ringgit
and the Chevron stock. The cumulants are plotted as a function of the asset weight
$w_1$, where the index $1$ corresponds to CHV, with the normalization $w_1 + w_2 = 1$.
Thus, the weight of the Malaysian Ringgit
is $w_2=1-w_1$. The error bars shown are obtained assuming an uncertainty 
in the determination of the exponents $\gamma_i = \gamma_i \pm 0.05$.}
\label{fig2}
\end{figure}

\begin{figure}
\epsfxsize=15cm
\epsfbox{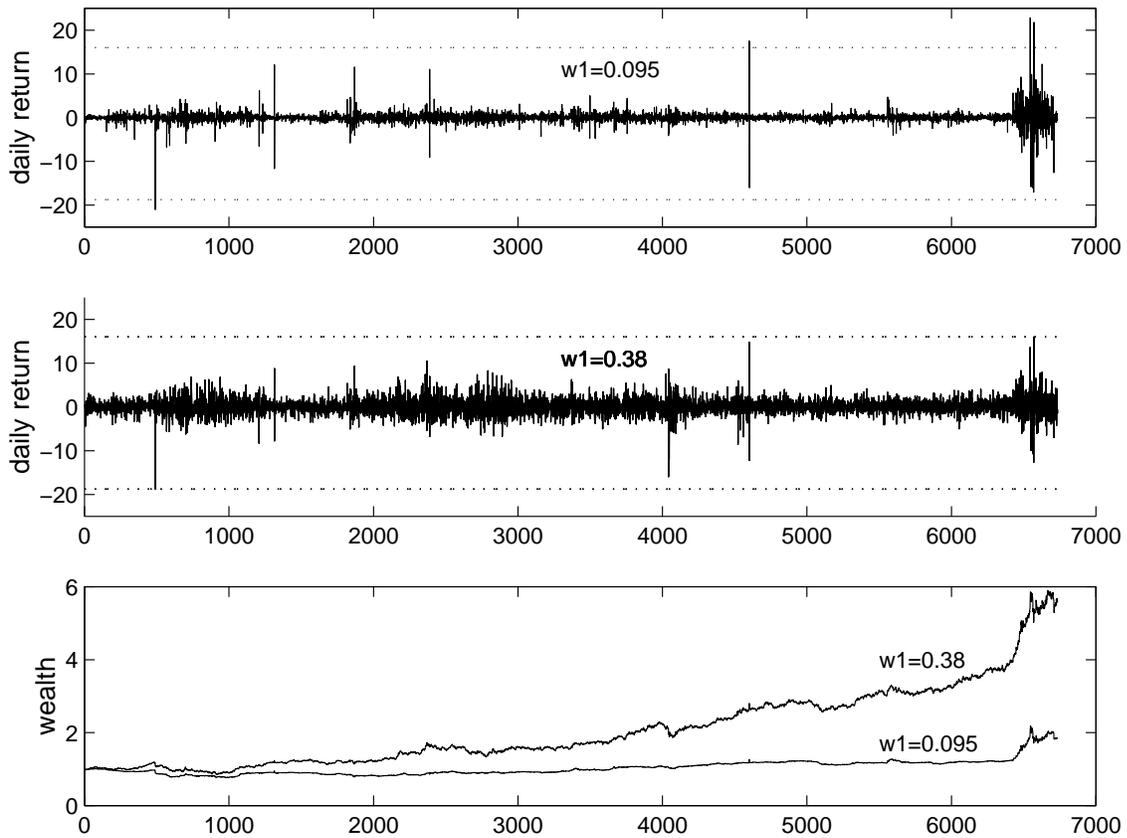}
\caption{Annualized daily returns (in percent)
 and cumulative wealth (starting with a unit wealth at time zero) 
 for the two portfolios corresponding to the
minimum variance with Chevron weight $w_1=0.095$ and  minimum fourth-order 
cumulant $c_4$ with Chevron weight $w_1=0.38$, determined from figure~\ref{fig2}. }
\label{fig3}
\end{figure}

Fig.~\ref{fig5} shows the efficient frontiers for 
portfolios constituted of the three assets CHV-XON-MYR. The lines are derived from the
theoretical prediction given by
Eq.~(\ref{optimalconditions}) using the exponents
determined from Fig.~\ref{fig1}. The solid line shows the mean-variance
efficient frontier normalized to the minimum variance and the 
dotted line shows the $c_1-c_4$ efficient frontier normalized to the minimum
fourth-order cumulant determined from the theory
assuming no correlations between the assets. The $+$ (resp. $o$) are the empirical mean-variance
(resp. $c_1-c_4$) portfolios constructed 
by scanning the weights $w_1$ (Chevron), $w_2$ (Exxon) and $w_3$ (Malaysian Ringgit) in the 
interval $[0,1]$ by steps of $0.02$ with the condition of normalization 
(\ref{sumomega}). Both family define a set of accessible portfolios and the frontier of each
domain define the corresponding empirical efficient frontiers. 
Note that by allowing negative weights (short position), 
the domains within the parabola are progressively filled up, corresponding to accessible portfolios
with ``short'' positions.

\begin{figure}
\epsfxsize=15cm
\epsfbox{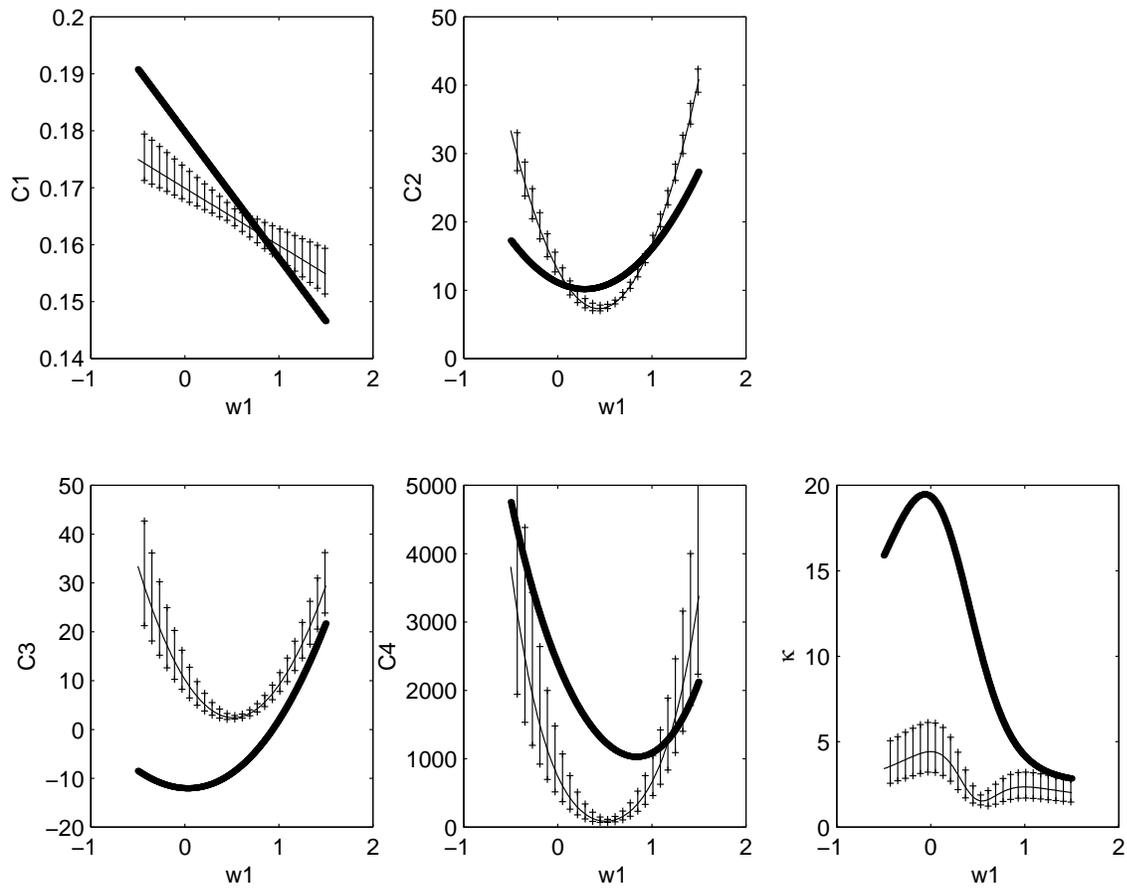}
\caption{Same as figure~\ref{fig2} for a portfolio constituted of the Exxon and the Chevron 
stocks. The cumulants are plotted as a function of the Chevron weight
$w_1$ and the weight of the Exxon stock is $w_2=1-w_1$.}
\label{fig4}
\end{figure}

\begin{figure}
\epsfxsize=15cm
\epsfbox{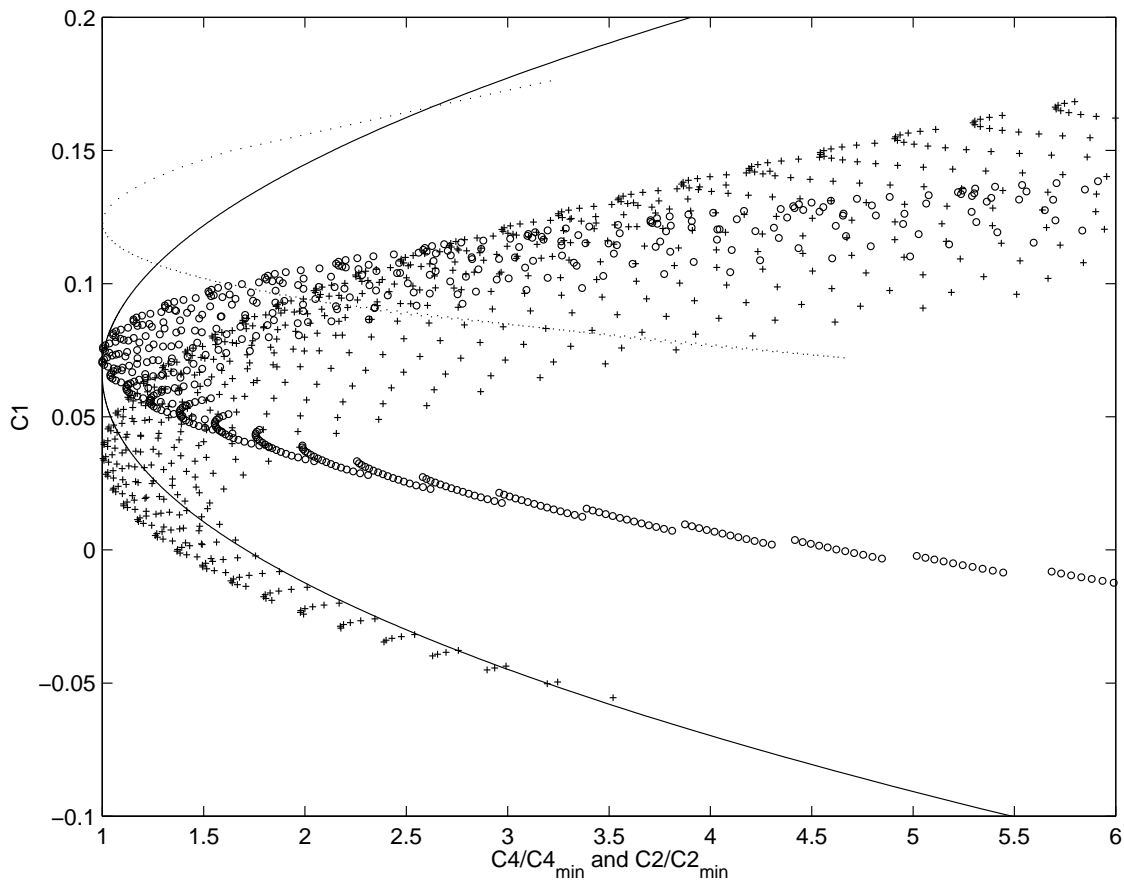}
\caption{Efficient frontiers for the three-asset portfolio CHV-XON-MYR derived from theory 
Eq.~(\ref{optimalconditions}) using the exponents $\gamma_i$'s
determined from Fig.~(\ref{fig1}). The solid line shows the mean-variance
efficient frontier normalized to the minimum variance and the 
dotted line shows the $c_1-c_4$ efficient frontier normalized to the minimum
fourth-order cumulant determined from the theory
assuming no correlations between the assets.  The $+$ (resp. $o$) are the empirical mean-variance
($c_1-c_4$) (resp. $c_1-c_4$) portfolios constructed 
by scanning the weights $w_1$ (Chevron), $w_2$ (Exxon) and $w_3$ (Malaysian Ringgit) in the 
interval $[0,1]$ by steps of $0.02$ while still implementing the condition of normalization 
(\ref{sumomega}). Both family define a set of accessible portfolios excluding
any ``short'' positions and the frontier of each
domain define the corresponding empirical frontiers. }
\label{fig5}
\end{figure}

The agreement is not good quantitatively between
theory and empirical tests due to the strong correlations between Chevron and Exxon which is
neglected in the theory (see figure ~(\ref{fig4})). However,
there is good qualitative agreement: the theory and empirical tests 
are essentially translated vertically, with the same characteristics.
The most important feature is that the $c_1-c_4$ portofolio with minimum fourth-order
cumulant (small ``large risks'') has a significantly larger return $c_1$ than the
portfolio with the minimum variable.
 For instance in the historical data, the return for the minimum variance occurs 
for $w_1=0.032, w_2=0.084, w_3=0.884$ for which the mean annualized return is $c_1=3.1\%$ and
the fourth-order cumulant is  $c_4/c_{4min}=2.22$, i.e. more than twice the minimum possible value.
The minimum of $c_4$ is reached for $w_1=0.292, w_2=0.084, w_3=0.624$ for 
which the mean annualized return is $c_1=7.2\%$, i.e. 
more than double the return for optimal the mean-variance
portfolio. Its variance is $c_2/c_{2min}=1.73$ which is a relatively moderate increase of 
``small risks''.  The results presented here can be easily generalized to higher cumulants
with similar conclusions.

\vskip 1cm

REFERENCES\,:

Artzner, P., F. Delbaen, J.M. Eber and D. Heath, 1996, A characterization of measures of
risks, Preprint 1996/14, Department of Mathematics, ULP Strasbourg.

Bouchaud, J.-P., D. Sornette, C. Walter and J.-P. Aguilar, 1998,
Taming large events: Optimal portfolio theory for strongly fluctuating assets, 
International Journal of Theoretical and Applied Finance 1, 25-41.

Embrechts, P., A.McNeil and D.Straumann, May 1999, Correlation: Pitfalls and Alterna-
tives, RISK 5, 69-71.
   
Embrechts, P., A.McNeil and D. Straumann, 1999, Correlation and dependency in Risk 
Management: properties and pitfalls, in  Proceedings of The Risk
Management Workshop, October 3, 1998 at The Newton Institute Cambridge, 
Cambridge University Press.
   
Frisch, U. and D. Sornette, 1997, Extreme deviations and applications,
Journal de Physique I France 7, 1155-1171.

Gopikrishnan, P., M. Meyer, L.A. Nunes Amaral and H.E. Stanley, 1998, 
Inverse cubic law for the distribution of stock price variations,
European Physical Journal B 3, 139-140.

Jorion, P., 1997, Value-at-Risk: The New Benchmark for
Controlling Derivatives Risk (Irwin Publishing, Chicago, IL).

Karlen, D., 1998, Using projections and correlations to approximate probability
distributions, Computer in Physics 12, 380-384.

Laherr\`ere, J. and D. Sornette, 1998,
Stretched exponential distributions in Nature and Economy: ``Fat tails''
with characteristic scales, European Physical Journal B 2, 525-539.

Litterman, R. and K. Winkelmann, 1998, Estimating covariance matrices (Risk Management
Series, Goldman Sachs).

Lux, T., 1996,
The stable Paretian hypothesis and the frequency of large returns: an
examination of major German stocks, Applied Financial Economics 6, n6, 463-475.

Lux, T. and M. Marchesi, 1999,
Scaling and criticality in a stochastic multi-agent model of a financial
market, Nature 397, 498-500.

Markovitz, H., 1959, Portfolio selection : Efficient diversification of
investments (John Wiley and Sons, New York).

Merton, R.C., 1990,  Continuous-time finance, (Blackwell, Cambridge).

Moran, P.A.P., 1969, Statistical inference with bivariate gamma distributions, 
Biometrika 56, 627-634.

Sornette, D., 1998, Large deviations and portfolio optimization, Physica A 256, 251-283.

Sornette, D., J. V. Andersen and P. Simonetti, 1998,
Minimizing volatility increases large risks, submitted to Risk Magazine,
preprint at http://xxx.lanl.gov/abs/cond-mat/9811292

Sornette, D., P. Simonetti and J. V. Andersen, 1999,
``Nonlinear'' covariance matrix and portfolio theory
for non-Gaussian multivariate distributions, submitted to The Review of
Financial Studies, preprint at
http://econwpa.wustl.edu/eprints/fin/papers/9902/9902004.abs

Stuart, A. and J.K. Ord, 1994,
Kendall's advanced theory of statistics,  6th ed. (London : Edward Arnold ;
New York : Halsted Press).

\pagebreak

 \end{document}